\documentclass[twocolumn]{aastex702}

\def \s{~\rm{s}}

\def \K{~\rm{K}}
\def \g{~\rm{g}}
\def \G{~\rm{G}}

\def \rmModot{~\rm{M_\odot}}

\def \rmModot{~\rm{M_{\sun}}}

\begin{document}

\title{The 2025.6 X-ray Minimum of Eta Carinae}

\shorttitle{Eta Car X-ray 2025.6}
\shortauthors{A. Kashi and N.Soker}

\author[0000-0002-7840-0181]{Amit Kashi} \thanks{E-mail: \href{mailto:kashi@ariel.ac.il}{kashi@ariel.ac.il}}
\email{kashi@ariel.ac.il}
\affiliation{Department of Physics, Ariel University, Ariel, 4070000, Israel}
\affiliation{Astrophysics, Geophysics, and Space Science (AGASS) Center, Ariel University, Ariel, 4070000, Israel}
\author[0000-0003-0375-8987]{Noam Soker} \thanks{E-mail: \href{mailto:soker@technion.ac.il}{soker@technion.ac.il}}
\email{soker@technion.ac.il}
\affiliation{Department of Physics, Technion - Israel Institute of Technology, Haifa, 3200003, Israel}
 
\begin{abstract}
We present observations of the recent 2025.6 X-ray minimum and recovery of $\eta$~Carinae, obtained with NICER/XTI, Swift/XRT, XRISM/Xtend, and Chandra/HRC-I. We construct light curves in the soft (0.5--2 keV) and hard (2--10 keV) bands and derive hardness ratios. The X-ray emission reached its minimum about 10 days earlier than in the previous five cycles. As in previous events, the spectrum softened during the decline, while the X-ray luminosity varied on short timescales. These findings strengthen the conclusion that enhanced line-of-sight absorption, such as that expected during an eclipse, does not have a significant role in producing the spectroscopic event. Instead, they indicate that the X-ray decline is caused by stochastic variations and instabilities in the two stellar winds and their interaction, destruction of the colliding winds region, and accretion of primary wind onto the secondary. The main process behind the deep X-ray minimum near periastron is the accretion of mass lost by the primary star onto the secondary star. Accretion at periastron passages in $\eta$ Carinae implies that accretion occurred at much higher rates during the nineteenth-century Great and Lesser Eruptions, when the primary star lost mass at much higher rates. Such accretion likely launches jets. The broader implication is that accretion events in binary systems, possibly with jet launching, power other luminous blue variable giant eruptions and intermediate-luminosity optical transients.
\end{abstract}



\section{INTRODUCTION}
\label{sec:intro}

Eta Carinae is a nearby, exceptionally luminous massive binary whose behavior is governed by the interaction of two powerful stellar winds \citep{DavidsonHumphreys2012}. The system follows a highly eccentric ($e\simeq0.9$), 5.54-year orbit \citep{Damineli1997}. Its luminous-blue-variable primary drives a dense, relatively slow wind, whereas the hotter, visually obscured companion supplies ionizing radiation and a much faster wind \citep{Hillier2001,PittardCorcoran2002,Verner2005}. Their collision produces a curved, time-dependent wind--collision region that generates luminous thermal X-rays and accelerates gas to high velocities \citep{Henley2008,Okazaki2008,Parkin2009,Hamaguchi2014}.

The system is known for the giant eruptions it underwent in the nineteenth century \citep[e.g.][]{DavidsonHumphreys1997,SmithFrew2011}. In \citet{KashiSoker2010} we modeled the orbital evolution of $\eta$ Carinae during its Great Eruption (1837.9--1858) and Lesser Eruption, accounting for both mass loss and mass transfer. 
The model showed that the Lesser Eruption (1887.3--1895.3) onset closely aligned with a periastron passage, supporting the \cite{Soker2001JetEta} hypothesis that secondary interaction triggered the event. 
By assuming that the primary luminosity peaks of the Great Eruption were also synchronized with periastron passages, we calculated the evolution of the orbit, taking a fraction of the mass lost by the primary during the Great Eruption to be accreted by the secondary. 
The gravitational energy released via this accretion process accounts for the radiated energy of the eruptions, while outflows or jets from the accreting companion may have shaped the bipolar Homunculus nebula. 
The orbital solution we obtained in \citet{KashiSoker2010} favors high stellar masses: a pre-Great Eruption total binary mass of $M_1+M_2 \gtrsim 250~\rmModot$, yielding post-eruptions masses of $M_1\simeq170$--$200~\rmModot$ and $M_2\simeq60$--$80~\rmModot$. 
This mass estimate was subsequently supported by independent line observations, where we argued that such an elevated secondary mass is a necessary consequence of supplying the extra Great Eruption luminosity through accretion while satisfying the observed orbital dynamics \citep{KashiSoker2016}.
The star’s apparent UV and visual brightness increased by several magnitudes from 2000 due to a smaller rate of dust formation in the ejecta \citep{Davidsonetal2018}. \cite{Davidsonetal2024} suggested that the star returned to the pre-Great Eruption state, though with much smaller mass.

\citet{Kashietal2021} analyzed the X-ray behavior of the 2020.1 periastron passage and associated X-ray minimum using NICER observations.
The analysis focused primarily on the harder energy band $2$--$10~\rm{keV}$ because its emission originates mainly in the shocked secondary wind near the apex of the colliding winds region and therefore provides a direct probe of the absorption toward the central binary.
The observations showed, as in previous cycles, the pre-periastron rise and flaring, followed by a deep non-zero minimum and recovery. The broad-band minimum lasted $\simeq25$--$37$ days (depending on the adopted criterion), whereas the hard-band minimum lasted about $23$ days. Comparison with the preceding four monitored cycles indicated that the 2020 event exhibited the steepest and earliest recovery.
\cite{EspinozaGaleasetal2022} also presented observations of the 2020.1 periastron passage and suggested that the X-ray minimum is caused by a combination of an eclipse and the disruption of the shock around the secondary of $\eta$ Car by the wind of the primary.

In this work, we present new X-ray observations of the 2025.6 periastron passage, compare them to previous events and the visible observations reported by \cite{Daminelietal2026}, and discuss their implications. In section \ref{sec:xray_obs} we describe the data reduction process of the new observations. In section \ref{sec:results} we present the resulting light curve and compare it to previous cycles.
In section \ref{sec:Discussion} we summarize and discuss our results in the context of luminous blue variable (LBV) eruptions and other transients.

\section{X-ray observations across the 2025.6 minimum}
\label{sec:xray_obs}

We constructed the X-ray light curves of $\eta$~Car from observations obtained by NICER/XTI, \textit{Swift}/XRT, XRISM/Xtend, and \textit{Chandra}/HRC-I. We defined the soft and hard bands as 0.5--2.0~keV and 2.0--10.0~keV whenever the detector provided useful pulse-height information. 
These bands were selected to distinguish the comparatively steady soft emission, which includes spatially extended emission, from the strongly variable hard emission associated primarily with the colliding winds region.
We assigned observation times to the exposure-weighted midpoint of the accepted good-time intervals (GTIs). We calculated rates and their uncertainties independently for each observation, with no interpolation between visits.
Table \ref{table:obs} summarizes the observations obtained for this work.

\begin{deluxetable*}{lcccccl}
\tablecaption{Summary of the $\eta$ Carinae X-Ray Observations\label{tab:xray_obs}}
\tabletypesize{\scriptsize}
\tablewidth{0pt}
\tablehead{
\colhead{Facility/} &
\colhead{Date Range} &
\colhead{$N_{\rm obs}$} &
\colhead{Observing Mode} &
\colhead{Energy Bands} &
\colhead{Exposure Range} &
\colhead{Normalization/} \\
\colhead{Instrument} &
\colhead{} &
\colhead{} &
\colhead{} &
\colhead{(keV)} &
\colhead{(s)} &
\colhead{Source Region}
}
\startdata
NICER/XTI      & 2024 Feb 24--2025 Jun 16 & 41 & Event & 0.5--2, 2--10 & 32--1,285 & Rates normalized to 52 FPMs \\
Swift/XRT      & 2025 Jul 19--2026 Jan 20 & 49 & PC and WT & 0.5--2, 2--10 & 178--1,758 & $20$-pixel ($47\farcs2$) radius \\
XRISM/Xtend    & 2024 Jun 11--2025 Sep 24 & 6  & Imaging & 0.5--2, 2--10 & 29,425--385,574 & $2\farcm5$ radius \\
Chandra/HRC-I  & 2025 May 29--2025 Sep 01 & 4  & Imaging & Broad band & 9,807--14,235 & $20$-pixel ($2\farcs64$) radius
\enddata
\tablecomments{FPM: focal-plane module. PC: photon counting. WT: windowed timing.}
\label{table:obs}
\end{deluxetable*}

\begin{figure*}
\centering
\includegraphics[trim= 1.0cm 1.0cm 0.0cm 0.0cm,clip=true,width=0.99\textwidth]{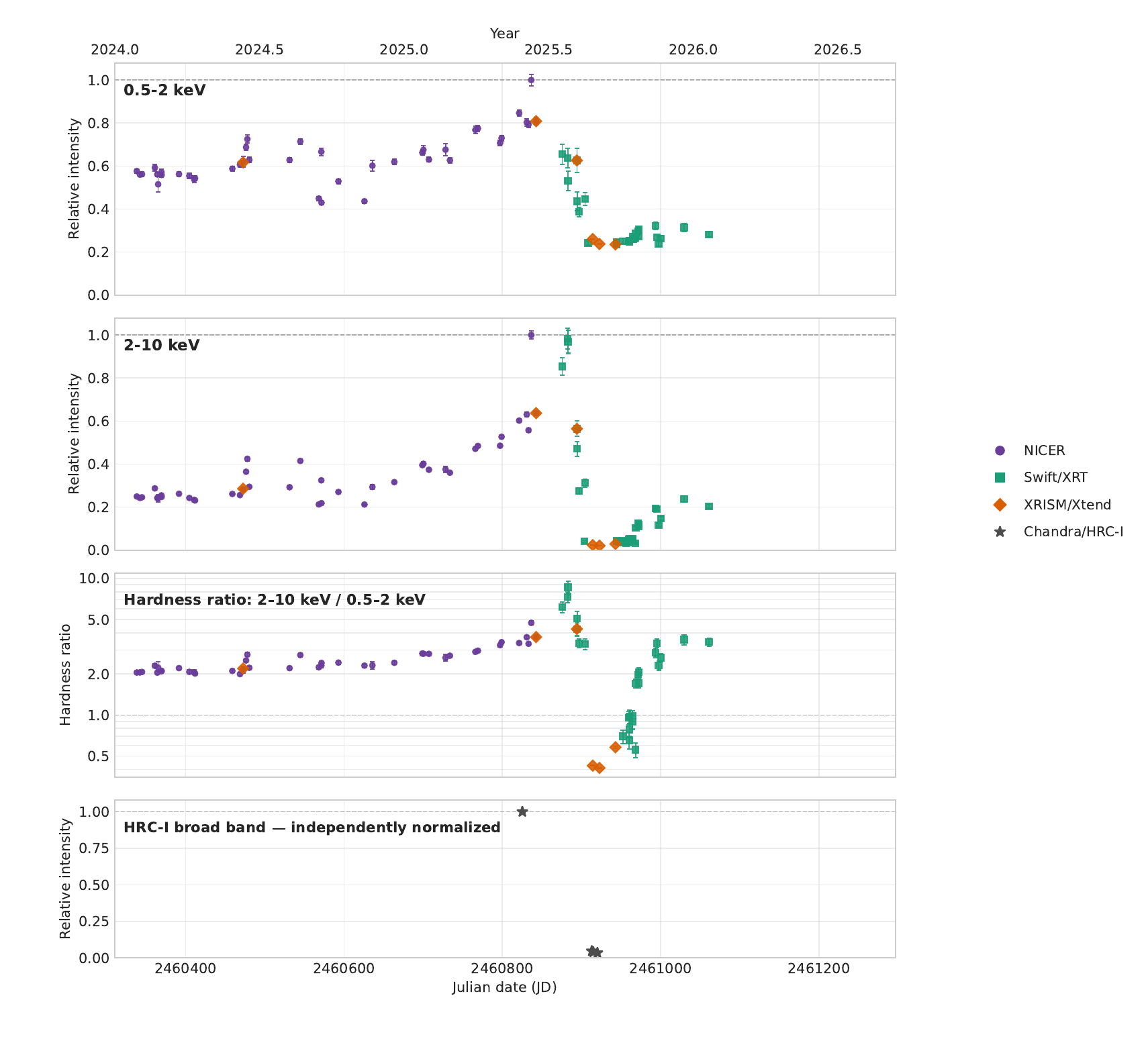} 
\caption{The X-ray light curve of $\eta$ Carinae across the 2025.6 minimum. From top to bottom, the panels show the soft component, hard component, the hardness ratio, and Chandra broad-band observations. The hard component is then collected into the light curve shown in Figures \ref{fig:cycles} and \ref{fig:V_X_ray} below. 
}
\label{fig:obsXray}
\end{figure*}

Because the four instruments have different effective areas, PSFs, apertures, and spectral responses, cross-mission comparisons are restricted to the timing and relative variability amplitude, and hardness ratios were calculated only from two bands measured by the same instrument.
Figure \ref{fig:obsXray} shows the light curve from all observations across the 2025.6 X-ray minimum. In the following, we describe how the observations were obtained from each of the instruments.
\begin{figure*}
\centering
\includegraphics[trim= 0.2cm 0.0cm 0.0cm 0.0cm,clip=true,width=0.99\textwidth]{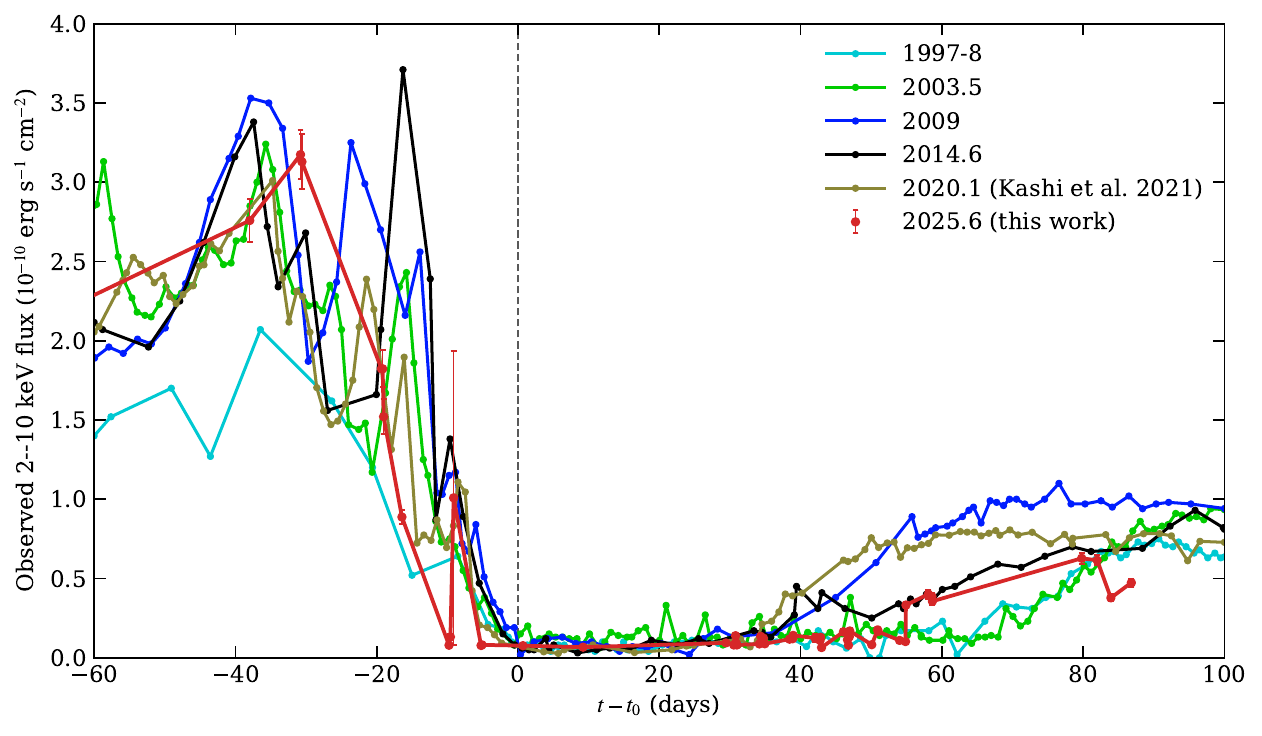} 
\caption{The X-ray light curve of $\eta$ Carinae across periastron passages over the last six cycles, showing the minima. The data from 2025.6 is new from this work. The data for the 2020.1 cycle are from \cite{Kashietal2021}, and the other cycles are adopted from \cite{Corcoranetal2017} and references therein. The dashed vertical line marks periastron passage (phase 0). 
}
\label{fig:cycles}
\end{figure*}

\subsection{NICER}
NICER provides non-imaging, high-throughput monitoring over approximately
0.2--12 keV \citep{2016SPIE.9905E..1HG}. We retrieved all public
observations of $\eta$ Carinae from HEASARC, up to the last available one on June 16, 2025, which did not cover the X-ray minimum but did show the early rise before. We processed each observation segment with the standard NICERDAS/HEASoft calibration and screening tools.
The soft and hard rates were extracted from PI channels 50--199 and 200--999, respectively. For each GTI,
the gross rate was calculated as $N/T$, with a Poisson uncertainty $\sqrt{N}/T$.
Both quantities were then converted to 52 FPM equivalent values using the detector normalization information appropriate to each GTI. Counts and exposures were summed over the accepted GTIs before deriving each observation level rate.
To remove variations caused solely by the number of enabled detectors, rates were scaled to a 52-detector equivalent by multiplying by $52/N_{\rm det}$. We report gross NICER rates rather than model-dependent background-subtracted fluxes. Thus, particularly in the soft band, the measurements include the approximately constant instrumental, diffuse Carina nebula, and spatially unresolved circumstellar contributions within the NICER field of view.

\subsection{Swift/XRT}
The XRT data \citep{2005SSRv..120..165B} were reduced separately for photon-counting (PC) and windowed-timing (WT) modes using standard HEASoft/\texttt{xrtpipeline} processing and Swift CALDB. Source and background products were generated following
the procedures of \citet{2009MNRAS.397.1177E}. For unpiled PC observations, source events were extracted from a 20 pixel radius circle ($47\farcs2$), while the background was measured in a source free 50 pixel circle on the same CCD. WT backgrounds were taken from an offset region along the readout strip. Individual snapshots were inspected for hot columns, field source contamination, and pile up. Whenever the PC rate exceeded the nominal pile up threshold, the PSF core was excluded until the outer radial profile was consistent with the calibrated PSF. The resulting rates were corrected for the excluded PSF fraction, vignetting, bad columns, and finite aperture with the exposure map based XRT correction. Background subtracted rates were then measured in 0.5--2.0 and 2.0--10.0 keV. Snapshot products within a single observation identifier were combined by summing source and area scaled background counts and exposures, thereby preserving Poisson counting statistics rather than averaging snapshot rates. Measurements below $3\sigma$ were omitted.

\subsection{XRISM/Xtend}
For the XRISM observations, we used the imaging CCD data from Xtend \citep{2025PASJ...77S...1T,2022SPIE12181E..1TM}. The archived data were reprocessed from unfiltered events with the mission pipeline \texttt{xapipeline}, using the calibration database current at the time of reduction, and the cleaned Xtend event files were subjected to the recommended GTI and event grade screening. Images were examined for flickering or anomalous pixels and for intervals of enhanced particle or scattered solar background. Affected detector pixels and time intervals were removed before photometry. Source counts were extracted from an aperture centered on $\eta$ Carinae, and the background was estimated from a source free region on the same CCD, avoiding the bright diffuse structures of the Carina Nebula. We checked each exposure for CCD pile up by comparing the central surface brightness profile and event grade distribution with the calibrated Xtend PSF. The 0.5--2.0 and 2.0--10.0~keV light curves were extracted with \texttt{XSELECT}, background subtracted after area scaling, and corrected for GTIs, bad pixels, and vignetting. Rates from separate observations were not added because the purpose of the Xtend measurements was to sample the approach to periastron and recovery from periastron.

\subsection{Chandra/HRC-I}
The archive has four HRC-I observations in the relevant time range. The observations were reprocessed with \texttt{chandra repro} in CIAO using the contemporaneous Chandra CALDB \citep{2002PASP..114....1W, 2006SPIE.6270E..1VF}. The reprocessing applied the standard HRC gain, degap, tap-ringing, aspect, GTI, and dead time corrections. Source events were extracted with \texttt{dmextract} from a circular aperture of 20 HRC pixels ($2\farcs64$) centered on the point source. The local background was measured in a concentric annulus of 50--100 pixels ($6\farcs6$--$13\farcs2$), after masking unrelated point sources, and was scaled by the ratio of geometrical areas before subtraction. The compact aperture minimizes contamination from the spatially extended outer ejecta, although unresolved emission within the central few arcseconds remains included. Owing to the very limited intrinsic energy resolution of HRC-I, reliable 0.5--2.0 and 2.0--10.0~keV rates cannot be derived from these data. We therefore retain one broad band, dead time corrected HRC-I count rate per observation and do not convert it to an energy flux. The HRC-I measurements are used only as high angular resolution checks on the temporal behavior measured with the spectrally resolving instruments.

\section{Results}
\label{sec:results}

\begin{figure*}
\centering
\includegraphics[trim= 0.0cm 0.7cm 0.0cm 0.0cm,clip=true,width=0.99\textwidth]{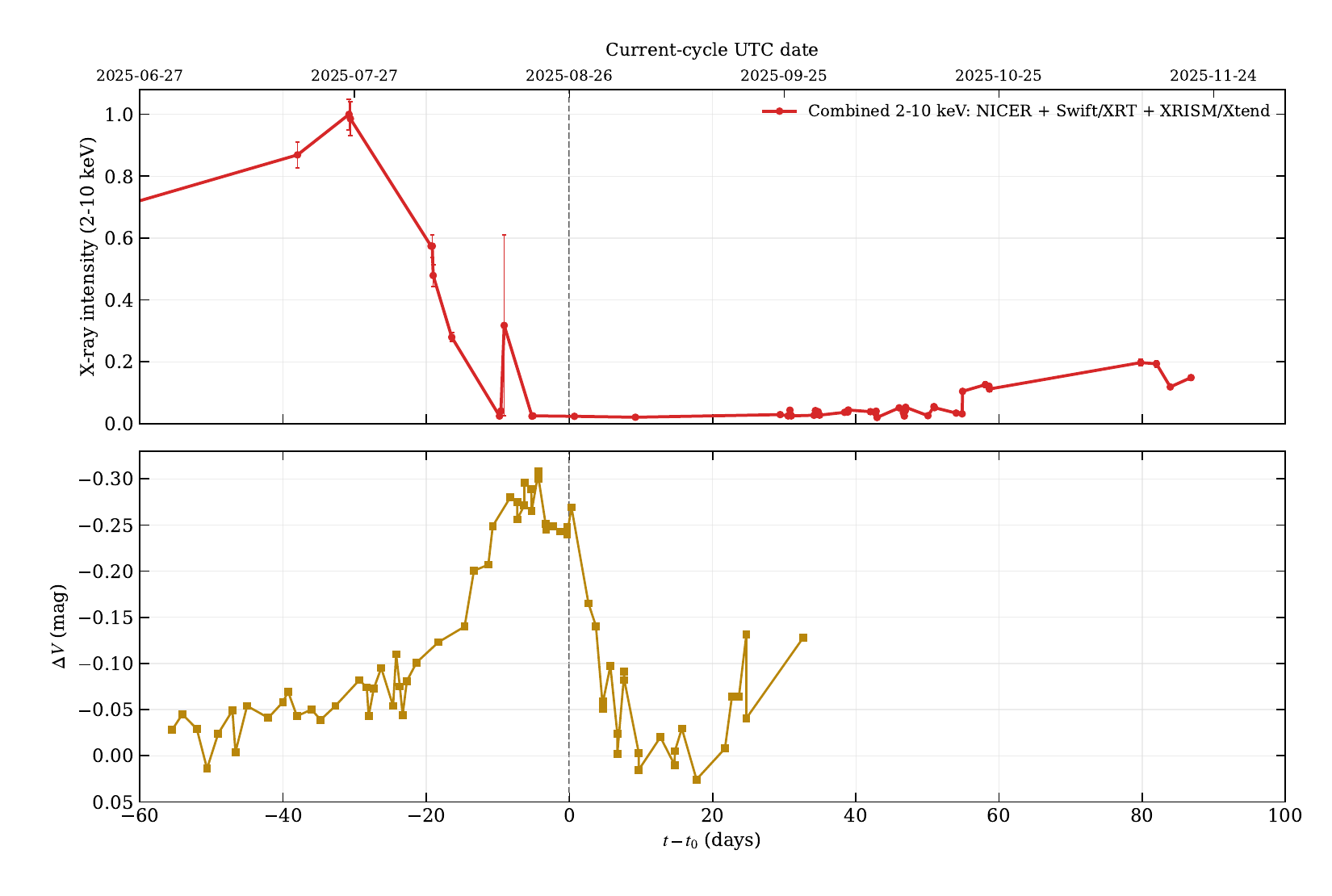} 
\caption{Comparison of the normalized 2025.6 event X-ray light curve of $\eta$ Car (this work) and the normalized V band light curve \citep{Daminelietal2026}.
}
\label{fig:V_X_ray}
\end{figure*}

\cite{Daminelietal2026} compares the normalized $V$-band light curves of $\eta$~Carinae during events across several orbital cycles, including the 2025.6 event (their figure~5). A recurrent photometric peak appears approximately 20 days before periastron, followed by a minimum centered near 5.2 days after, which they consider the time of superior conjunction. The peak amplitude decreased from $\Delta V\simeq0.45$ mag in 2003.5 and 2009 to $\simeq0.3$ mag in 2020.1 and 2025.6.
In Figure \ref{fig:V_X_ray} we plot the observations across the 2025.6 event, V band from \cite{Daminelietal2026}, and X-ray observations from this work.

\cite{Daminelietal2026} noted that the secondary luminosity is too low for an eclipse to explain the decrease in the luminosity near periastron. The decrease between apastron and periastron is  $\Delta V \simeq 0.4$, which requires that $\simeq 30\%$  of the central binary luminosity is absorbed. 
They suggested that the increase in luminosity before periastron passage is due to reflection of light by material that is lost by the primary, and that this material is eclipsed near periastron. They suggested that this material continues to reflect light along $\simeq 60\%$ of the orbit. This requires a quantitative demonstration, as it is not clear how this material stays close enough to the primary. 

We consider three observational properties of the last two spectroscopic events (X-ray minima), of 2020.1 and 2025.6:
\begin{enumerate}
    \item 
    These two events further strengthen the accuracy of the binary period and the softening of the X-ray emission as it declines. As many studies have noted (e.g., \citealt{AkashiSokerBehar2006, Kashietal2021, EspinozaGaleasetal2022}), the softening of the X-ray emission implies that the main cause of the X-ray minimum is not an eclipse or absorbing material, but the collapse of the colliding winds region, i.e., the secondary star accretes mass near periastron passages (e.g., \citealt{AkashiSokerBehar2006, AkashiKashiSoker2013, Kashi2017, Daminelietal2026}). The X-ray decline and the softening coincide (Figure \ref{fig:obsXray}), indicating that absorption plays a very small role. This hints that the dense primary wind does not absorb the colliding winds region, i.e., the secondary star is closer to us near periastron passages (e.g., \citealt{AkashiSokerBehar2006, Kashietal2021}).   
    \item The visible light curve of \cite{Daminelietal2026}, their Figure 5, shows that while in four events the visible minimum is at $t\simeq 5$ days post-periastron, the minimum of the 2020.2 event is about a week later. This suggests the minimum is not due to an exact binary phase, such as an eclipse, but rather to the wind properties near periastron passage.   
    \item The new X-ray light curve towards the minimum of 2025.6 that we present in Figure \ref{fig:cycles}, alongside the preceding five events, shows that in the 2025.6 event the system reached the X-ray minimum $\simeq 10$ days earlier. This again shows that the two stars' winds and the accretion process dominate the light curve, not absorption. This supports the claim that the eclipse, if it exists, plays a minor role. 
    We note the X-ray peak at $t \simeq 9$ days before periastron in all events; however, in the last event it might be less significant (uncertainties are large). The peak occurs after the light curve has already declined steeply. This steep decline results from the destruction of the colliding winds region and the beginning of accretion onto the secondary star \citep[e.g][]{AkashiSokerBehar2006}. We attribute the X-ray peak at $t\approx -9$ days to a high accretion rate onto the secondary star as the wind-colliding material is accreted. This emission results from the colliding winds region impacting the secondary, as simulations by \cite{Kashi2017} show, particularly for the high mass model for the two stars. 
\end{enumerate}

Overall, the last two cycles support the notion that the properties of the two winds, their interaction, and accretion onto the secondary star determine the light curve. The large fluctuations in the X-ray emission towards the minimum and in the visible along the entire orbit, with large variation between cycles (e.g., \citealt{Daminelietal2026}), also show that the winds' properties, with their stochastic variations and instabilities, determine the light curve. In other words, \textit{there are no indications of an eclipse near periastron passages}. The eclipse is not a complete one in this model, as the orbit is inclined and we observe the system at $i=41^\circ$ from the equatorial plane of $\eta$ Carinae.  
Although these arguments do not rule out the possibility that the secondary is farther away near periastron (as claimed by, e.g., \citealt{Nielsenetal2007, Madura2013, Weigelt2016, Strawnetal2023, XrismCollaborationetal2026, Daminelietal2026}), they are better aligned with the view that the secondary star is closer to us near periastron passage (e.g., \citealt{Falceta-Goncalvesetal2005,KashiSoker2007b, KashiSoker2008,Kashietal2021}).

\section{Discussion and summary}
\label{sec:Discussion}

We presented the X-ray light curve of the 2025.6 spectroscopic event of $\eta$ Carinae in two wave bands and their ratio (hardness ratio) in Figure \ref{fig:obsXray}. In Figure \ref{fig:cycles}, we compare it with those of five earlier events. 
The X-ray emission in the last event declined to its minimum value about 10 days earlier than in the five earlier events (note that the observations are sparse so it might have reached the minimum even 16 days earlier). In Figure \ref{fig:V_X_ray} we compare the X-ray light curve to the visible lightcurve. The event occurred as expected, showing that the binary system maintains its period. 

The last two spectroscopic events strengthen earlier findings and notions (Section \ref{sec:results}). We note these three properties: (1) As in all previous events, the X-ray emission becomes softer as the X-ray declines. This indicates that the decline is not due to absorption. (2) The minimum in the visible of the previous event, 2020.1, was a bout a week later than those of earlier events and the last one \citep{Daminelietal2026}. This suggests the minimum is not set by an exact superior conjunction of the secondary star, which would make all minima occur at the same time. (3) The X-ray minimum of the last event started $\simeq 10$~days earlier than those of earlier events. This again hints that the minimum is not caused by an exact superior conjunction of the secondary.  

The above three findings, and the continuous fluctuations of the visible and X-ray emission, support the notion that the spectroscopic event is set by the properties of the two winds and their interaction, including accretion of the primary wind onto the secondary star (e.g., \citealt{Soker2005AccEtaCar, AkashiSokerBehar2006, KashiSoker2007a, AkashiKashiSoker2013, Kashi2017, Daminelietal2026}).
We find no indication of the occurrence of a substantial extra absorption, or, in other words, the secondary star is likely to be closer to us near periastron passages, as \cite{Kashietal2021} showed quantitatively by calculating the column density to the X-ray emitting gas. 
The last event further supports the claim that the primary process of the spectroscopic event is mass accretion by the secondary star: instead of blowing a wind, the secondary star accretes mass from the primary wind, as simulations show (e.g., \citealt{Kashi2017}). 

The X-ray minimum in the past six cycle lasted 23--70 days \citep{Corcoranetal2017, Kashietal2021}, while simulations showed that the accretion phase lasts for about 20--60 days, depending on the adopted parameters \citep{Kashi2017}.
The understanding that accretion onto the secondary star is the primary process during the spectroscopic event, and is actually responsible for it, has much broader implications. In all the studied spectroscopic events, the primary star of $\eta$ Carinae was calm. During the Great Eruption and the Lesser Eruption of (e.g., \citealt{DavidsonHumphreys1997, Humphreysetal1999}), the primary was more active and lost mass at much higher rates. This dense wind prolongs the accretion phase and leads to much more energetic events. Namely, the behavior of the spectroscopic event strongly supports the claim that the \textit{primary process of giant eruptions of LBVs is  accretion onto a companion} \citep[e.g.,][]{KashiSoker2010,Kashi2010,MukhijaKashi2024,MukhijaKashi2026,BearSoker2025}.

The Great Eruption was also explained by models involving a tripple system \citep{PortegiesZwartvandenHeuvel2016, Hirai2021}. However, as we noted earlier (e.g., \citealt{Soker2024, BearSoker2025RAA}), these models face difficulties, and we consider them unlikely. The main reasons are:
(1) The Lesser Eruption required another merging star under the triple star scenario. Namely, a system of four initial stars. (2) The present binary system and the Homunculus share an equatorial plane \citep[e.g.,[]{Maduraetal2012}, implying a coplanar triple stellar system; merger scenarios require an unstable triple system that is unlikely to be coplanar. (3) The triple star scenario of \cite{Hirai2021} results in the presence of a dense gas in the equatorial plane; this is not observed in the Homunculus. For these difficulties, we prefer the accretion onto the secondary scenario in a binary system.

Two decades ago, the main question about major eruptions of LBVs was whether they are single-star eruptions (e.g., \citealt{Smith2007ASPC}) or whether binary interaction powers them (e.g., \citealt{Bath1979, Notaetal1995}). Nowadays, a key question remains whether the companion launches jets that power LBV eruptions. 
\cite{Bath1979} suggested that mass accretion onto a main-sequence star powered the $\eta$ Carinae Great Eruption, and even mentioned winds from the accretion disk. \cite{Soker2001JetEta} proposed that jets launched by the secondary star played the primary role in shaping the ejecta and powering the Great and Lesser Eruptions of $\eta$ Carinae; we quantified the accretion process and jet launching \citep{KashiSoker2010}. In earlier papers, we gave accretion in a binary system a primary role in powering other intermediate-luminosity optical transients (ILOTs; \citealt{Kashietal2010,KashiSoker2010arXiv}), a heterogeneous group that includes LBVs, luminous red novae, some pre-supernova eruptions, and more (e.g., \citealt{Blagorodnova2026Proc, Fraser2026Proc,KaminskiBlagorodnova2026}). This process is known as the High Accretion Powered ILOT (HAPI; \citealt{SokerKashi2016}) model.

As mentioned, a major question in the study of giant LBV eruptions and ILOTs is the role of jets in powering them.
While some works do not consider jets important (e.g., \citealt{ChenIvanova2024, Kirilovetal2025, ChenZhuo2026, HutchinsonSmithetal2026, SarinHirai2026, Sneppenetal2026, Twumetal2026}, and references to many earlier papers therein), others give jets the primary role in powering all ILOTs, including cases where the system enters common-envelope evolution (e.g., \citealt[e.g.,][]{KashiSoker2016, SokerKashi2016, SokerKaplan2021, Soker2024}). Support for the major role of jets comes from the bipolar structure of several ILOTs (e.g., \citealt{Kaminskietal2018V4332, Kaminskietal2021CKVul, Kaminski2024, Chesneauetal2014, Kaminskietal2021V838Mon, Mobeenetal2024}). 
Some energetic LBVs require the main sequence companion to accrete mass at very high rates. Recent simulations show that when the jets that the accretion disk launches remove high entropy accreted gas, the accretion rate can be very large: the jets operate in a positive feedback cycle (as they facilitate their own strong power), in what is termed ``jetted mass removal accretion scenario''.  
The jetted mass removal accretion scenario can operate in accretion onto massive main sequence stars (\citealt{BearSoker2025, Scolnicetal2025}), and onto stripped-envelope stars, like Wolf-Rayet (WR) stars \citep{CohenBearSoker2026}. 
  
The accreting secondary star might also be a neutron star. \cite{GilkisSokerKashi2019} raised the possibility that a neutron star companion on an eccentric orbit that launched pairs of jets powered the LBV pre-supernova eruptions of SN 2009ip. \cite{Aghakhanlooetal2026} suggested a similar model for the LBV AT 2016blu.

To summarize, the new X-ray light curve of the 2025.6 spectroscopic event of $\eta$ Carinae adds a small amount of support to (1) the orientation model where the secondary star is closer to us at periastron, and (2) that accretion onto the companion is the main driver of the spectroscopic event, including the X-ray decline. The accretion process at periastron passages, supported by many other arguments, further supports accretion in a binary system during the Great and Lesser Eruption of $\eta$ Carinae in the nineteenth century, when mass loss from the primary star was much larger. More generally, these conclusions are compatible with the notion that LBVs and many other ILOTs are powered by accretion that most likely launches jets. 

\vspace{0.5cm}

\section*{Acknowledgments}
We acknowledge the Ariel HPC Center at Ariel University for providing computing resources that have contributed to the research results reported within this paper.
This research was performed using resources from Ariel Visualization Laboratory.
NS acknowledges a grant from the Pazy Foundation 2026, which supported this research, and the Charles Wolfson Academic Chair at the Technion.


\bibliography{refs}{}

@ARTICLE{Kashietal2021,
       author = {{Kashi}, Amit and {Principe}, David A. and
                 {Soker}, Noam and {Kastner}, Joel H.},
        title = "{The X-Ray Properties of Eta Carinae during Its 2020 X-Ray Minimum}",
      journal = {\apj},
         year = 2021,
        month = jun,
       volume = {914},
       number = {1},
          eid = {47},
        pages = {47},
          doi = {10.3847/1538-4357/abfa9c},
archivePrefix = {arXiv},
       eprint = {2010.03877},
 primaryClass = {astro-ph.HE},
       adsurl = {https://ui.adsabs.harvard.edu/abs/2021ApJ...914...47K}
}

@INPROCEEDINGS{2016SPIE.9905E..1HG,
       author = {{Gendreau}, Keith C. and {Arzoumanian}, Zaven and {Adkins}, Phillip W. and {Albert}, Cheryl L. and {Anders}, John F. and {Aylward}, Andrew T. and {Baker}, Charles L. and {Balsamo}, Erin R. and {Bamford}, William A. and {Benegalrao}, Suyog S. and {Berry}, Daniel L. and {Bhalwani}, Shiraz and {Black}, J. Kevin and {Blaurock}, Carl and {Bronke}, Ginger M. and {Brown}, Gary L. and {Budinoff}, Jason G. and {Cantwell}, Jeffrey D. and {Cazeau}, Thoniel and {Chen}, Philip T. and {Clement}, Thomas G. and {Colangelo}, Andrew T. and {Coleman}, Jerry S. and {Coopersmith}, Jonathan D. and {Dehaven}, William E. and {Doty}, John P. and {Egan}, Mark D. and {Enoto}, Teruaki and {Fan}, Terry W. and {Ferro}, Deneen M. and {Foster}, Richard and {Galassi}, Nicholas M. and {Gallo}, Luis D. and {Green}, Chris M. and {Grosh}, Dave and {Ha}, Kong Q. and {Hasouneh}, Monther A. and {Heefner}, Kristofer B. and {Hestnes}, Phyllis and {Hoge}, Lisa J. and {Jacobs}, Tawanda M. and {J{\o}rgensen}, John L. and {Kaiser}, Michael A. and {Kellogg}, James W. and {Kenyon}, Steven J. and {Koenecke}, Richard G. and {Kozon}, Robert P. and {LaMarr}, Beverly and {Lambertson}, Mike D. and {Larson}, Anne M. and {Lentine}, Steven and {Lewis}, Jesse H. and {Lilly}, Michael G. and {Liu}, Kuochia Alice and {Malonis}, Andrew and {Manthripragada}, Sridhar S. and {Markwardt}, Craig B. and {Matonak}, Bryan D. and {Mcginnis}, Isaac E. and {Miller}, Roger L. and {Mitchell}, Alissa L. and {Mitchell}, Jason W. and {Mohammed}, Jelila S. and {Monroe}, Charles A. and {Montt de Garcia}, Kristina M. and {Mul{\'e}}, Peter D. and {Nagao}, Louis T. and {Ngo}, Son N. and {Norris}, Eric D. and {Norwood}, Dwight A. and {Novotka}, Joseph and {Okajima}, Takashi and {Olsen}, Lawrence G. and {Onyeachu}, Chimaobi O. and {Orosco}, Henry Y. and {Peterson}, Jacqualine R. and {Pevear}, Kristina N. and {Pham}, Karen K. and {Pollard}, Sue E. and {Pope}, John S. and {Powers}, Daniel F. and {Powers}, Charles E. and {Price}, Samuel R. and {Prigozhin}, Gregory Y. and {Ramirez}, Julian B. and {Reid}, Winston J. and {Remillard}, Ronald A. and {Rogstad}, Eric M. and {Rosecrans}, Glenn P. and {Rowe}, John N. and {Sager}, Jennifer A. and {Sanders}, Claude A. and {Savadkin}, Bruce and {Saylor}, Maxine R. and {Schaeffer}, Alexander F. and {Schweiss}, Nancy S. and {Semper}, Sean R. and {Serlemitsos}, Peter J. and {Shackelford}, Larry V. and {Soong}, Yang and {Struebel}, Jonathan and {Vezie}, Michael L. and {Villasenor}, Joel S. and {Winternitz}, Luke B. and {Wofford}, George I. and {Wright}, Michael R. and {Yang}, Mike Y. and {Yu}, Wayne H.},
        title = "{The Neutron star Interior Composition Explorer (NICER): design and development}",
    booktitle = {Space Telescopes and Instrumentation 2016: Ultraviolet to Gamma Ray},
         year = 2016,
       editor = {{den Herder}, Jan-Willem A. and {Takahashi}, Tadayuki and {Bautz}, Marshall},
       series = {Society of Photo-Optical Instrumentation Engineers (SPIE) Conference Series},
       volume = {9905},
        month = jul,
          eid = {99051H},
        pages = {99051H},
          doi = {10.1117/12.2231304},
       adsurl = {https://ui.adsabs.harvard.edu/abs/2016SPIE.9905E..1HG}
}

@ARTICLE{EspinozaGaleasetal2022,
       author = {{Espinoza-Galeas}, David and {Corcoran}, M.~F. and {Hamaguchi}, K. and {Russell}, C.~M.~P. and {Gull}, T.~R. and {Moffat}, A.~F.~J. and {Richardson}, N.~D. and {Weigelt}, G. and {Hillier}, D. John and {Damineli}, Augusto and {Stevens}, Ian R. and {Madura}, Thomas and {Gendreau}, K. and {Arzoumanian}, Z. and {Navarete}, Felipe},
        title = "{NICER X-Ray Observations of Eta Carinae during Its Most Recent Periastron Passage}",
      journal = {\apj},
         year = 2022,
        month = jul,
       volume = {933},
       number = {2},
          eid = {136},
        pages = {136},
          doi = {10.3847/1538-4357/ac69ce},
archivePrefix = {arXiv},
       eprint = {2207.03457},
 primaryClass = {astro-ph.HE},
       adsurl = {https://ui.adsabs.harvard.edu/abs/2022ApJ...933..136E}
}

@ARTICLE{2005SSRv..120..165B,
       author = {{Burrows}, David N. and {Hill}, J.~E. and {Nousek}, J.~A. and {Kennea}, J.~A. and {Wells}, A. and {Osborne}, J.~P. and {Abbey}, A.~F. and {Beardmore}, A. and {Mukerjee}, K. and {Short}, A.~D.~T. and {Chincarini}, G. and {Campana}, S. and {Citterio}, O. and {Moretti}, A. and {Pagani}, C. and {Tagliaferri}, G. and {Giommi}, P. and {Capalbi}, M. and {Tamburelli}, F. and {Angelini}, L. and {Cusumano}, G. and {Br{\"a}uninger}, H.~W. and {Burkert}, W. and {Hartner}, G.~D.},
        title = "{The Swift X-Ray Telescope}",
      journal = {\ssr},
         year = 2005,
        month = oct,
       volume = {120},
       number = {3-4},
        pages = {165-195},
          doi = {10.1007/s11214-005-5097-2},
archivePrefix = {arXiv},
       eprint = {astro-ph/0508071},
 primaryClass = {astro-ph},
       adsurl = {https://ui.adsabs.harvard.edu/abs/2005SSRv..120..165B}
}

@ARTICLE{2009MNRAS.397.1177E,
       author = {{Evans}, P.~A. and {Beardmore}, A.~P. and {Page}, K.~L. and {Osborne}, J.~P. and {O'Brien}, P.~T. and {Willingale}, R. and {Starling}, R.~L.~C. and {Burrows}, D.~N. and {Godet}, O. and {Vetere}, L. and {Racusin}, J. and {Goad}, M.~R. and {Wiersema}, K. and {Angelini}, L. and {Capalbi}, M. and {Chincarini}, G. and {Gehrels}, N. and {Kennea}, J.~A. and {Margutti}, R. and {Morris}, D.~C. and {Mountford}, C.~J. and {Pagani}, C. and {Perri}, M. and {Romano}, P. and {Tanvir}, N.},
        title = "{Methods and results of an automatic analysis of a complete sample of Swift-XRT observations of GRBs}",
      journal = {\mnras},
         year = 2009,
        month = aug,
       volume = {397},
       number = {3},
        pages = {1177-1201},
          doi = {10.1111/j.1365-2966.2009.14913.x},
archivePrefix = {arXiv},
       eprint = {0812.3662},
 primaryClass = {astro-ph},
       adsurl = {https://ui.adsabs.harvard.edu/abs/2009MNRAS.397.1177E}
}

@ARTICLE{2025PASJ...77S...1T,
       author = {{Tashiro}, Makoto and {Kelley}, Richard and {Watanabe}, Shin and {Maejima}, Hironori and {Reichenthal}, Lillian and {Toda}, Kenichi and {Hartz}, Leslie and {Santovincenzo}, Andrea and {Matsushita}, Kyoko and {Yamaguchi}, Hiroya and {Petre}, Robert and {Williams}, Brian and {Guainazzi}, Matteo and {Costantini}, Elisa and {Takei}, Yoh and {Ishisaki}, Yoshitaka and {Fujimoto}, Ryuichi and {Henegar-Leon}, Joy and {Sneiderman}, Gary and {Tomida}, Hiroshi and {Mori}, Koji and {Nakajima}, Hiroshi and {Terada}, Yukikatsu and {Holland}, Matthew and {Loewenstein}, Michael and {Miller}, Eric and {Sawada}, Makoto and {Kallman}, Timothy and {Kaastra}, Jelle and {Done}, Chris and {Enoto}, Teruaki and {Bamba}, Aya and {Corrales}, Lia and {Ueda}, Yoshihiro and {Kara}, Erin and {Zhuravleva}, Irina and {Fujita}, Yutaka and {Arai}, Yoshitaka and {Audard}, Marc and {Awaki}, Hisamitsu and {Ballhausen}, Ralf and {Baluta}, Chris and {Bando}, Nobutaka and {Behar}, Ehud and {Bialas}, Thomas and {Boissay-Malaquin}, Rozenn and {Brenneman}, Laura and {Brown}, Gregory V. and {Chiao}, Meng and {Cumbee}, Renata and {de Vries}, Cor and {den Herder}, Jan-Willem and {D{\'\i}az Trigo}, Mar{\'\i}a and {DiPirro}, Michael and {Dotani}, Tadayasu and {Carrero}, Jacobo Ebrero and {Ebisawa}, Ken and {Eckart}, Megan and {Eckert}, Dominique and {Eguchi}, Satoshi and {Ezoe}, Yuichiro and {Ferrigno}, Carlo and {Foster}, Adam and {Fukazawa}, Yasushi and {Fukushima}, Kotaro and {Furuzawa}, Akihiro and {Gallo}, Luigi C. and {Garcia Martinez}, Javier and {Gorter}, Nathalie and {Grim}, Martin and {Gu}, Liyi and {Hagino}, Kouichi and {Hamaguchi}, Kenji and {Hatsukade}, Isamu and {Hayashi}, Katsuhiro and {Hayashi}, Takayuki and {Hell}, Natalie and {Hodges-Kluck}, Edmund and {Horiuchi}, Takafumi and {Hornschemeier}, Ann and {Hoshino}, Akio and {Ichinohe}, Yuto and {Ikuta}, Chisato and {Iizuka}, Ryo and {Ishi}, Daiki and {Ishida}, Manabu and {Ishihama}, Naoki and {Ishikawa}, Kumi and {Ishimura}, Kosei and {Jaffe}, Tess and {Katsuda}, Satoru and {Kanemaru}, Yoshiaki and {Kenyon}, Steven and {Kilbourne}, Caroline and {Kimball}, Mark and {Kitamoto}, Shunji and {Kobayashi}, Shogo and {Kohmura}, Takayoshi and {Kubota}, Aya and {Leutenegger}, Maurice A. and {Maeda}, Yoshitomo and {Markevitch}, Maxim and {Matsumoto}, Hironori and {Matsuzaki}, Keiichi and {McCammon}, Dan and {McLaughlin}, Brian and {McNamara}, Brian and {Mernier}, Fran{\c{c}}ois and {Miko}, Joseph and {Miller}, Jon M. and {Minesugi}, Kenji and {Mitani}, Shinji and {Mitsuishi}, Ikuyuki and {Mizumoto}, Misaki and {Mizuno}, Tsunefumi and {Mukai}, Koji and {Murakami}, Hiroshi and {Mushotzky}, Richard and {Nakazawa}, Kazuhiro and {Natsukari}, Chikara and {Ness}, Jan-Uwe and {Nigo}, Kenichiro and {Nishiyama}, Mari and {Nobukawa}, Kumiko and {Nobukawa}, Masayoshi and {Noda}, Hirofumi and {Odaka}, Hirokazu and {Ogawa}, Mina and {Ogawa}, Shoji and {Ogorzalek}, Anna and {Okajima}, Takashi and {Okamoto}, Atsushi and {Ota}, Naomi and {Ozaki}, Masanobu and {Paltani}, Stephane and {Plucinsky}, Paul and {Porter}, F. Scott and {Pottschmidt}, Katja and {Quero}, Jose Antonio and {Sasaki}, Takahiro and {Sato}, Kosuke and {Sato}, Rie and {Sato}, Toshiki and {Sato}, Yoichi and {Seta}, Hiromi and {Shida}, Maki and {Shidatsu}, Megumi and {Shigeto}, Shuhei and {Shipman}, Russel and {Shinozaki}, Keisuke and {Shirron}, Peter and {Simionescu}, Aurora and {Smith}, Randall K. and {Soong}, Yang and {Suzuki}, Hiromasa and {Szymkowiak}, Andrew and {Takahashi}, Hiromitsu and {Takeo}, Mai and {Tamagawa}, Toru and {Tamura}, Keisuke and {Tanaka}, Takaaki and {Tanimoto}, Atsushi and {Terashima}, Yuichi and {Tsuboi}, Yohko and {Tsujimoto}, Masahiro and {Tsunemi}, Hiroshi and {Tsuru}, Takeshi Go and {Uchida}, Hiroyuki and {Uchida}, Nagomi and {Uchida}, Yuusuke and {Uchiyama}, Hideki and {Uno}, Shinichiro and {Vink}, Jacco and {Witthoeft}, Michael and {Wolfs}, Rob and {Yamada}, Satoshi and {Yamada}, Shinya and {Yamaoka}, Kazutaka and {Yamasaki}, Noriko and {Yamauchi}, Makoto and {Yamauchi}, Shigeo and {Yanagase}, Keiichi and {Yaqoob}, Tahir and {Yasuda}, Susumu and {Yoneyama}, Tomokage and {Yoshida}, Tessei and {Yukita}, Mihoko},
        title = "{X-Ray Imaging and Spectroscopy Mission}",
      journal = {\pasj},
         year = 2025,
        month = sep,
       volume = {77},
        pages = {S1-S9},
          doi = {10.1093/pasj/psaf023},
       adsurl = {https://ui.adsabs.harvard.edu/abs/2025PASJ...77S...1T}
}

@INPROCEEDINGS{2022SPIE12181E..1TM,
       author = {{Mori}, Koji and {Tomida}, Hiroshi and {Nakajima}, Hiroshi and {Okajima}, Takashi and {Noda}, Hirofumi and {Tanaka}, Takaaki and {Uchida}, Hiroyuki and {Hagino}, Kouichi and {Kobayashi}, Shogo Benjamin and {Suzuki}, Hiromasa and {Yoshida}, Tessei and {Murakami}, Hiroshi and {Uchiyama}, Hideki and {Nobukawa}, Masayoshi and {Nobukawa}, Kumiko and {Yoneyama}, Tomokage and {Matsumoto}, Hironori and {Tsuru}, Takeshi and {Yamauchi}, Makoto and {Hatsukade}, Isamu and {Ishida}, Manabu and {Maeda}, Yoshitomo and {Hayashi}, Takayuki and {Tamura}, Keisuke and {Boissay-Malaquin}, Rozenn and {Sato}, Toshiki and {Hiraga}, Junko and {Kohmura}, Takayoshi and {Yamaoka}, Kazutaka and {Dotani}, Tadayasu and {Ozaki}, Masanobu and {Tsunemi}, Hiroshi and {Kanemaru}, Yoshiaki and {Sato}, Jin and {Takaki}, Toshiyuki and {Terada}, Yuta and {Miyazaki}, Keitaro and {Kusunoki}, Kohei and {Otsuka}, Yoshinori and {Yokosu}, Haruhiko and {Yonemaru}, Wakana and {Asahina}, Yoh and {Asakura}, Kazunori and {Yoshimoto}, Marina and {Ode}, Yuichi and {Sato}, Junya and {Hakamata}, Tomohiro and {Aoyagi}, Mio and {Aoki}, Yuma and {Tsunomachi}, Shun and {Doi}, Toshiki and {Aoki}, Daiki and {Fujisawa}, Kaito and {Kitajima}, Masatoshi and {Hayashida}, Kiyoshi},
        title = "{Xtend, the soft x-ray imaging telescope for the X-Ray Imaging and Spectroscopy Mission (XRISM)}",
    booktitle = {Space Telescopes and Instrumentation 2022: Ultraviolet to Gamma Ray},
         year = 2022,
       editor = {{den Herder}, Jan-Willem A. and {Nikzad}, Shouleh and {Nakazawa}, Kazuhiro},
       series = {Society of Photo-Optical Instrumentation Engineers (SPIE) Conference Series},
       volume = {12181},
        month = aug,
          eid = {121811T},
        pages = {121811T},
          doi = {10.1117/12.2626894},
archivePrefix = {arXiv},
       eprint = {2303.07575},
 primaryClass = {astro-ph.IM},
       adsurl = {https://ui.adsabs.harvard.edu/abs/2022SPIE12181E..1TM}
}

@ARTICLE{2002PASP..114....1W,
       author = {{Weisskopf}, M.~C. and {Brinkman}, B. and {Canizares}, C. and {Garmire}, G. and {Murray}, S. and {Van Speybroeck}, L.~P.},
        title = "{An Overview of the Performance and Scientific Results from the Chandra X-Ray Observatory}",
      journal = {\pasp},
         year = 2002,
        month = jan,
       volume = {114},
       number = {791},
        pages = {1-24},
          doi = {10.1086/338108},
archivePrefix = {arXiv},
       eprint = {astro-ph/0110308},
 primaryClass = {astro-ph},
       adsurl = {https://ui.adsabs.harvard.edu/abs/2002PASP..114....1W}
}

@INPROCEEDINGS{2006SPIE.6270E..1VF,
       author = {{Fruscione}, Antonella and {McDowell}, Jonathan C. and {Allen}, Glenn E. and {Brickhouse}, Nancy S. and {Burke}, Douglas J. and {Davis}, John E. and {Durham}, Nick and {Elvis}, Martin and {Galle}, Elizabeth C. and {Harris}, Daniel E. and {Huenemoerder}, David P. and {Houck}, John C. and {Ishibashi}, Bish and {Karovska}, Margarita and {Nicastro}, Fabrizio and {Noble}, Michael S. and {Nowak}, Michael A. and {Primini}, Frank A. and {Siemiginowska}, Aneta and {Smith}, Randall K. and {Wise}, Michael},
        title = "{CIAO: Chandra's data analysis system}",
    booktitle = {Observatory Operations: Strategies, Processes, and Systems},
         year = 2006,
       editor = {{Silva}, David R. and {Doxsey}, Rodger E.},
       series = {Society of Photo-Optical Instrumentation Engineers (SPIE) Conference Series},
       volume = {6270},
        month = jun,
          eid = {62701V},
        pages = {62701V},
          doi = {10.1117/12.671760},
       adsurl = {https://ui.adsabs.harvard.edu/abs/2006SPIE.6270E..1VF}
}

@ARTICLE{Damineli1997,
       author = {{Damineli}, Augusto and {Conti}, Peter S. and {Lopes}, Dalton F.},
        title = "{Eta Carinae: a long period binary?}",
      journal = {New Astronomy},
         year = 1997,
       volume = {2},
        pages = {107--117},
          doi = {10.1016/S1384-1076(97)00008-0},
       adsurl = {https://ui.adsabs.harvard.edu/abs/1997NewA....2..107D}
}

@ARTICLE{Hillier2001,
       author = {{Hillier}, D. John and {Davidson}, Kris and
                 {Ishibashi}, Kazunori and {Gull}, Theodore},
        title = "{On the Nature of the Central Source in Eta Carinae}",
      journal = {\apj},
         year = 2001,
       volume = {553},
        pages = {837--860},
          doi = {10.1086/320948},
       adsurl = {https://ui.adsabs.harvard.edu/abs/2001ApJ...553..837H}
}

@ARTICLE{PittardCorcoran2002,
       author = {{Pittard}, Julian M. and {Corcoran}, Michael F.},
        title = "{In Hot Pursuit of the Hidden Companion of Eta Carinae:
                  An X-Ray Determination of the Wind Parameters}",
      journal = {\aap},
         year = 2002,
       volume = {383},
        pages = {636--647},
          doi = {10.1051/0004-6361:20020025},
       adsurl = {https://ui.adsabs.harvard.edu/abs/2002A&A...383..636P}
}

@ARTICLE{Verner2005,
       author = {{Verner}, E. and {Bruhweiler}, F. and {Gull}, T.},
        title = "{The Binarity of Eta Carinae Revealed from Photoionization
                  Modeling of the Spectral Variability of the Weigelt
                  Blobs B and D}",
      journal = {\apj},
         year = 2005,
       volume = {624},
        pages = {973--982},
          doi = {10.1086/429400},
       adsurl = {https://ui.adsabs.harvard.edu/abs/2005ApJ...624..973V}
}

@ARTICLE{Henley2008,
       author = {{Henley}, D.~B. and {Corcoran}, M.~F. and
                 {Pittard}, J.~M. and {Stevens}, I.~R. and
                 {Hamaguchi}, K. and {Gull}, T.~R.},
        title = "{Chandra X-Ray Grating Spectrometry of Eta Carinae near
                  X-Ray Minimum. I. Variability of the Sulfur and Silicon
                  Emission Lines}",
      journal = {\apj},
         year = 2008,
       volume = {680},
        pages = {705--727},
          doi = {10.1086/587472},
       adsurl = {https://ui.adsabs.harvard.edu/abs/2008ApJ...680..705H}
}

@ARTICLE{Okazaki2008,
       author = {{Okazaki}, Atsuo T. and {Owocki}, Stanley P. and
                 {Russell}, Christopher M.~P. and {Corcoran}, Michael F.},
        title = "{Modelling the RXTE Light Curve of Eta Carinae from a
                  3D SPH Simulation of Its Binary Wind Collision}",
      journal = {\mnras},
         year = 2008,
       volume = {388},
        pages = {L39--L43},
          doi = {10.1111/j.1745-3933.2008.00496.x},
       adsurl = {https://ui.adsabs.harvard.edu/abs/2008MNRAS.388L..39O}
}

@ARTICLE{Parkin2009,
       author = {{Parkin}, E.~R. and {Pittard}, J.~M. and
                 {Corcoran}, M.~F. and {Hamaguchi}, K. and
                 {Stevens}, I.~R.},
        title = "{3D Modelling of the Colliding Winds in Eta Carinae:
                  Evidence for Radiative Inhibition}",
      journal = {\mnras},
         year = 2009,
       volume = {394},
        pages = {1758--1774},
          doi = {10.1111/j.1365-2966.2009.14475.x},
       adsurl = {https://ui.adsabs.harvard.edu/abs/2009MNRAS.394.1758P}
}

@ARTICLE{Hamaguchi2014,
       author = {{Hamaguchi}, Kenji and {Corcoran}, Michael F. and
                 {Russell}, Christopher M.~P. and {Pollock}, A.~M.~T. and
                 {Gull}, Theodore R. and {Teodoro}, Mairan and
                 {Madura}, Thomas I. and {Damineli}, Augusto and
                 {Pittard}, Julian M.},
        title = "{X-Ray Emission from Eta Carinae near Periastron in 2009.
                  I. A Two-State Solution}",
      journal = {\apj},
         year = 2014,
       volume = {784},
          eid = {125},
        pages = {125},
          doi = {10.1088/0004-637X/784/2/125},
archivePrefix = {arXiv},
       eprint = {1401.5870},
 primaryClass = {astro-ph.SR},
       adsurl = {https://ui.adsabs.harvard.edu/abs/2014ApJ...784..125H}
}

@ARTICLE{Madura2013,
       author = {{Madura}, Thomas I. and {Gull}, Theodore R. and
                 {Okazaki}, Atsuo T. and {Russell}, Christopher M.~P. and
                 {Owocki}, Stanley P. and {Groh}, Jose H. and
                 {Corcoran}, Michael F. and {Hamaguchi}, Kenji and
                 {Teodoro}, Mairan},
        title = "{Constraints on Decreases in Eta Carinae's Mass-Loss from
                  3D Hydrodynamic Simulations of Its Binary Colliding Winds}",
      journal = {\mnras},
         year = 2013,
       volume = {436},
        pages = {3820--3855},
          doi = {10.1093/mnras/stt1871},
       adsurl = {https://ui.adsabs.harvard.edu/abs/2013MNRAS.436.3820M}
}

@ARTICLE{Weigelt2016,
       author = {{Weigelt}, G. and {Hofmann}, K.-H. and {Schertl}, D. and
                 {Clementel}, N. and {Corcoran}, M.~F. and {Damineli}, A. and
                 {de Wit}, W.-J. and {Grellmann}, R. and {Groh}, J. and
                 {Guieu}, S. and {Gull}, T. and {Heininger}, M. and
                 {Hillier}, D.~J. and {Hummel}, C.~A. and {Kraus}, S. and
                 {Madura}, T. and {Mehner}, A. and {M{\'e}rand}, A. and
                 {Millour}, F. and {Moffat}, A.~F.~J. and {Ohnaka}, K. and
                 {Patru}, F. and {Petrov}, R.~G. and {Rengaswamy}, S. and
                 {Richardson}, N.~D. and {Rivinius}, T. and
                 {Sch{\"o}ller}, M. and {Teodoro}, M. and {Wittkowski}, M.},
        title = "{VLTI-AMBER Velocity-resolved Aperture-synthesis Imaging
                  of Eta Carinae with a Spectral Resolution of 12,000}",
      journal = {\aap},
         year = 2016,
       volume = {594},
          eid = {A106},
        pages = {A106},
          doi = {10.1051/0004-6361/201628832},
       adsurl = {https://ui.adsabs.harvard.edu/abs/2016A&A...594A.106W}
}

@ARTICLE{Daminelietal2026,
       author = {{Damineli}, Augusto and {Almeida}, L.~A. and
                 {Jablonski}, Francisco J. and {Fern{\'a}ndez-Laj{\'u}s}, Eduardo and
                 {Navarete}, Felipe and {Martioli}, Eder and
                 {Weigelt}, Gerd and {Capobiango}, Rodrigo},
        title = "{Eta Carinae's historical light curve: evidence for cyclic Roche lobe overflow from the primary star}",
      journal = {arXiv e-prints},
         year = 2026,
        month = aug,
          eid = {arXiv:2608.16818},
        pages = {arXiv:2608.16818},
          doi = {10.48550/arXiv.2608.16818},
archivePrefix = {arXiv},
       eprint = {2608.16818},
 primaryClass = {astro-ph.SR},
       adsurl = {https://ui.adsabs.harvard.edu/abs/2026arXiv260816818D}
}

@PROCEEDINGS{DavidsonHumphreys2012,
        title = "{Eta Carinae and the Supernova Impostors}",
    booktitle = {Eta Carinae and the Supernova Impostors},
         year = 2012,
       editor = {{Davidson}, Kris and {Humphreys}, Roberta M.},
       series = {Astrophysics and Space Science Library},
       volume = {384},
        month = jan,
          doi = {10.1007/978-1-4614-2275-4},
       adsurl = {https://ui.adsabs.harvard.edu/abs/2012ASSL..384.....D}
}

@ARTICLE{DavidsonHumphreys1997,
       author = {{Davidson}, Kris and {Humphreys}, Roberta M.},
        title = "{Eta Carinae and Its Environment}",
      journal = {\araa},
         year = 1997,
        month = jan,
       volume = {35},
        pages = {1-32},
          doi = {10.1146/annurev.astro.35.1.1},
       adsurl = {https://ui.adsabs.harvard.edu/abs/1997ARA&A..35....1D}
}

@INPROCEEDINGS{Davidsonetal2024,
       author = {{Davidson}, K. and {Martin}, J.~C. and {Ishibashi}, K. and {Humphreys}, R.~M.},
        title = "{Eta Carinae Firmly Established in a New State}",
    booktitle = {American Astronomical Society Meeting Abstracts \#244},
         year = 2024,
       series = {American Astronomical Society Meeting Abstracts},
       volume = {244},
        month = jun,
          eid = {310.04},
        pages = {310.04},
       adsurl = {https://ui.adsabs.harvard.edu/abs/2024AAS...24431004D}
}

@ARTICLE{Davidsonetal2018,
       author = {{Davidson}, Kris and {Ishibashi}, Kazunori and {Martin}, John C. and {Humphreys}, Roberta M.},
        title = "{Eta Carinae{\textquoteright}s Declining Outflow Seen in the UV, 2002-2015}",
      journal = {\apj},
         year = 2018,
        month = may,
       volume = {858},
       number = {2},
          eid = {109},
        pages = {109},
          doi = {10.3847/1538-4357/aabdef},
archivePrefix = {arXiv},
       eprint = {1804.07400},
 primaryClass = {astro-ph.SR},
       adsurl = {https://ui.adsabs.harvard.edu/abs/2018ApJ...858..109D}
}

@ARTICLE{Kashi2017,
       author = {{Kashi}, Amit},
        title = "{Accretion at the periastron passage of Eta Carinae}",
      journal = {\mnras},
         year = 2017,
        month = jan,
       volume = {464},
       number = {1},
        pages = {775-782},
          doi = {10.1093/mnras/stw2303},
archivePrefix = {arXiv},
       eprint = {1609.03135},
 primaryClass = {astro-ph.SR},
       adsurl = {https://ui.adsabs.harvard.edu/abs/2017MNRAS.464..775K}
}

@ARTICLE{MukhijaKashi2024,
       author = {{Mukhija}, Bhawna and {Kashi}, Amit},
        title = "{Giant Eruptions in Massive Stars and their Effect on the Stellar Structure}",
      journal = {\apj},
         year = 2024,
        month = oct,
       volume = {974},
       number = {1},
          eid = {124},
        pages = {124},
          doi = {10.3847/1538-4357/ad7398},
archivePrefix = {arXiv},
       eprint = {2408.01718},
 primaryClass = {astro-ph.SR},
       adsurl = {https://ui.adsabs.harvard.edu/abs/2024ApJ...974..124M}
}

@ARTICLE{AkashiSokerBehar2006,
       author = {{Akashi}, Muhammad and {Soker}, Noam and {Behar}, Ehud},
        title = "{Accretion onto the Companion of {\ensuremath{\eta}} Carinae during the Spectroscopic Event. II. X-Ray Emission Cycle}",
      journal = {\apj},
         year = 2006,
        month = jun,
       volume = {644},
       number = {1},
        pages = {451-463},
          doi = {10.1086/503317},
archivePrefix = {arXiv},
       eprint = {astro-ph/0509429},
 primaryClass = {astro-ph},
       adsurl = {https://ui.adsabs.harvard.edu/abs/2006ApJ...644..451A}
}

@ARTICLE{Soker2005AccEtaCar,
       author = {{Soker}, Noam},
        title = "{Accretion by the Secondary in {\ensuremath{\eta}} Carinae During the Spectroscopic Event. I. Flow Parameters}",
      journal = {\apj},
         year = 2005,
        month = dec,
       volume = {635},
       number = {1},
        pages = {540-546},
          doi = {10.1086/497389},
archivePrefix = {arXiv},
       eprint = {astro-ph/0505218},
 primaryClass = {astro-ph},
       adsurl = {https://ui.adsabs.harvard.edu/abs/2005ApJ...635..540S}
}

@ARTICLE{AkashiKashiSoker2013,
       author = {{Akashi}, Muhammad S. and {Kashi}, Amit and {Soker}, Noam},
        title = "{Accretion of dense clumps in the periastron passage of {\ensuremath{\eta}} Carinae}",
      journal = {\na},
         year = 2013,
        month = jan,
       volume = {18},
        pages = {23-30},
          doi = {10.1016/j.newast.2012.05.010},
       adsurl = {https://ui.adsabs.harvard.edu/abs/2013NewA...18...23A}
}

@ARTICLE{Nielsenetal2007,
       author = {{Nielsen}, K.~E. and {Corcoran}, M.~F. and {Gull}, T.~R. and {Hillier}, D.~J. and {Hamaguchi}, K. and {Ivarsson}, S. and {Lindler}, D.~J.},
        title = "{{\ensuremath{\eta}} Carinae across the 2003.5 Minimum: Spectroscopic Evidence for Massive Binary Interactions}",
      journal = {\apj},
         year = 2007,
        month = may,
       volume = {660},
       number = {1},
        pages = {669-686},
          doi = {10.1086/513006},
archivePrefix = {arXiv},
       eprint = {astro-ph/0701632},
 primaryClass = {astro-ph},
       adsurl = {https://ui.adsabs.harvard.edu/abs/2007ApJ...660..669N}
}

@ARTICLE{KashiSoker2007a,
       author = {{Kashi}, Amit and {Soker}, Noam},
        title = "{Modelling the radio light curve of {\ensuremath{\eta}} Carinae}",
      journal = {\mnras},
         year = 2007,
        month = jul,
       volume = {378},
       number = {4},
        pages = {1609-1618},
          doi = {10.1111/j.1365-2966.2007.11908.x},
archivePrefix = {arXiv},
       eprint = {astro-ph/0702389},
 primaryClass = {astro-ph},
       adsurl = {https://ui.adsabs.harvard.edu/abs/2007MNRAS.378.1609K}
}

@ARTICLE{KashiSoker2007b,
       author = {{Kashi}, Amit and {Soker}, Noam},
        title = "{The source of the helium visible lines in {\ensuremath{\eta}} Carinae}",
      journal = {\na},
         year = 2007,
        month = oct,
       volume = {12},
       number = {7},
        pages = {590-596},
          doi = {10.1016/j.newast.2007.04.004},
archivePrefix = {arXiv},
       eprint = {astro-ph/0702661},
 primaryClass = {astro-ph},
       adsurl = {https://ui.adsabs.harvard.edu/abs/2007NewA...12..590K}
}

@ARTICLE{BearSoker2025RAA,
       author = {{Bear}, Ealeal and {Soker}, Noam},
        title = "{On the Response of Massive Main Sequence Stars to Mass Accretion and Outflow at High Rates}",
      journal = {Research in Astronomy and Astrophysics},
         year = 2025,
        month = feb,
       volume = {25},
       number = {2},
          eid = {025010},
        pages = {025010},
          doi = {10.1088/1674-4527/ada8ef},
archivePrefix = {arXiv},
       eprint = {2407.03182},
 primaryClass = {astro-ph.SR},
       adsurl = {https://ui.adsabs.harvard.edu/abs/2025RAA....25b5010B}
}

@ARTICLE{Soker2024,
       author = {{Soker}, Noam},
        title = "{More Luminous Red Novae That Require Jets}",
      journal = {Galaxies},
         year = 2024,
        month = jun,
       volume = {12},
       number = {4},
          eid = {33},
        pages = {33},
          doi = {10.3390/galaxies12040033},
archivePrefix = {arXiv},
       eprint = {2404.19617},
 primaryClass = {astro-ph.SR},
       adsurl = {https://ui.adsabs.harvard.edu/abs/2024Galax..12...33S}
}

@ARTICLE{Hirai2021,
       author = {{Hirai}, Ryosuke and {Podsiadlowski}, Philipp and {Owocki}, Stanley P. and {Schneider}, Fabian R.~N. and {Smith}, Nathan},
        title = "{Simulating the formation of {\ensuremath{\eta}} Carinae's surrounding nebula through unstable triple evolution and stellar merger-induced eruption}",
      journal = {\mnras},
         year = 2021,
        month = may,
       volume = {503},
       number = {3},
        pages = {4276-4296},
          doi = {10.1093/mnras/stab571},
archivePrefix = {arXiv},
       eprint = {2011.12434},
 primaryClass = {astro-ph.SR},
       adsurl = {https://ui.adsabs.harvard.edu/abs/2021MNRAS.503.4276H}
}
\bibliographystyle{aasjournalv7.1}



\end{document}